\documentclass[letterpaper]{article} 
\usepackage[margin=1.5in]{geometry}
\usepackage[hyphens]{url}  
\usepackage{graphicx} 
\usepackage[round]{natbib}
\usepackage{amsmath} 
\usepackage{amssymb}
\usepackage{float}
\usepackage{caption}
\usepackage{subcaption}
\usepackage{authblk}
\usepackage[bb=boondox,bbscaled=.95,cal=boondoxo]{mathalfa}
\title{``F*** You Biden'': \\ Cross-Partisan Electoral Toxicity on X}

\author{Danishjeet Singh\thanks{Email: singhdan@iu.edu}}
\author{Anindya Mondal}
\author{Filippo Menczer}

\affil{Observatory on Social Media, Indiana University, Bloomington, USA}
        
\date{\today}

\newcommand{\dataset}[1]{{\texttt{#1}}}
\newcommand{\twitterdataset}{\dataset{US2024-Twitter-Elections} }
\newcommand{\nposts}{261,501}
\newcommand{\nreplies}{2,415,781}
\newcommand{\ngold}{384}
\begin{document}
\maketitle

\begin{abstract}
Political discourse on social media has grown increasingly toxic, with electoral periods amplifying partisan hostility and cross-group attacks. 
Yet it remains unclear whether toxicity in online political speech reflects how partisans communicate within their own circles, or how aggressively they engage with the opposition. 
Disentangling these dynamics is critical for understanding online political hostility and for designing effective content moderation.
We examine this question at scale using a large collection of original posts and replies from X (formerly Twitter), collected during the 2024 U.S. presidential election. 
Using a human-validated large language model to classify the political alignment of posts and users, and the Perspective API for toxicity scoring, we uncover a striking asymmetry: Republican-leaning posts are significantly more toxic than Democratic-leaning posts, yet Democratic-leaning posts attract significantly more toxic replies.
To interpret this finding, we compare the toxicity of same-party and cross-partisan replies. 
While cross-partisan replies are slightly but significantly more toxic than same-party replies, this is true for both Democratic and Republican posts.  
However, Republican users account for a large majority of replies to Democratic posts, while Democrats account for a minority of replies to Republican content. 
Therefore, the elevated toxicity directed at Democratic content is better explained by the volume of Republican cross-partisan replies.
\end{abstract}

\section{Introduction}

A key mechanism is out-group engagement: negative content about the opposing party generates more interaction than positive content about one's own side \citep{Rathje2021}. While politicians tend to rally their base, ordinary users disproportionately engage with opposing content \citep{Yu2023}. Crucially, this cross-partisan exposure does not reduce polarization; it tends to amplify it \citep{Bail2018}.
Political incivility on social media is concentrated among a small but highly active and networked minority \citep{Rasmussen2022PoliticalRA}. Public figures are disproportionately targeted with personal attacks \citep{Belschner2023CritiqueOE}.

Yet the origins of reply toxicity, whether from co-partisans or the opposition, remain poorly understood. 
We address this gap by analyzing political posts and replies on X (formerly Twitter) during the 2024 U.S. presidential election. 
We find that Republican posts use more toxic language, yet Democratic posts attract more toxic replies, driven by the elevated toxicity of cross-partisan replies. 
The title of this paper is the most frequent phrase among highly toxic replies, illustrating such cross-partisan hostility. 

\section{Related Work}

Affective polarization, defined as partisan hostility rooted in social identity rather than policy disagreement, has intensified steadily in the United States with co-partisans increasingly socializing together and viewing the opposing party with contempt \citep{Iyengar2019}. 
On Twitter, this shapes network structure: retweet networks are strongly ideologically homophilic, while mention networks show more cross-partisan contact, particularly around breaking news events \citep{Truthy_icwsm2011politics,Barbera2015}.

Despite this segregation, cross-partisan engagement is both common and asymmetric \citep{Cetinkaya2025}. 
Posts about the out-group generate roughly four times the engagement of in-group content \citep{Rathje2021}, and ordinary users, unlike politicians, show stronger engagement with out-party attacks than in-party affirmation \citep{Yu2023}. 
This cross-partisan exposure does not reduce polarization: a field experiment found that Republicans who followed a liberal bot became more conservative \citep{Bail2018}, and a preregistered trial during the 2024 U.S. election confirmed that boosting user exposure to hostile partisan content on X increased affective polarization \citep{Piccardi2025}.

Politicians --- particularly incumbents and governing-party members --- are disproportionately targeted with personal attacks \citep{Belschner2023CritiqueOE}.
Analysis across nine countries finds that out-group interactions are consistently more toxic than in-group ones, but reveals no consistent left-right asymmetry in toxicity production \citep{Falkenberg2024}.

The U.S. context appears distinctive. 
Republican users have historically exhibited greater political activity and tighter social network cohesion on Twitter than Democrats \citep{Conover2012}. 
More recent work found that left-leaning Twitter users received significantly more toxic replies than right-leaning users in COVID-related discussions \citep{Xu2024}. 
In U.S. Twitter discussions in 2020, both Democratic and Republican accounts tended to post more toxic content in cross-partisan replies then same-party replies, but the difference was more marked for Democrats \citep{Cetinkaya2025}. 
These patterns motivate our present investigation into whether cross-partisan toxicity reflects structural differences in targeting behavior during elections.  

The present analysis relies on Large Language Models (LLMs) to perform scalable annotation of social media content. This reflect the broad application of such tools in computational social science (CSS). 
GPT-3.5 was shown to outperform crowd workers on text annotation tasks --- including political framing and stance detection --- at substantially lower cost \citep{Gilardi2023}. 
Systematic evaluations across dozens of CSS benchmarks confirmed that LLMs now match or exceed human-level performance on tasks that previously required prohibitive annotation labor \citep{Ziems2024}. 
For political classification specifically, LLMs effectively detect partisan ideology and political leanings in social media text without task-specific fine-tuning \citep{hernandes2024llmsleftrightcenter}. 

\section{Methods}

We utilize the \twitterdataset dataset \citep{twitterdataset}, comprising politically relevant Twitter content related to the 2024 U.S. Presidential Election, collected between May and November 2024 using a curated keyword list.
The dataset contains approximately 44 million tweets, including original posts, retweets, quotes, and replies.

For this study, we focus on original tweets and their direct replies. 
Quoted tweets are excluded due to ambiguity in attributing toxicity between the quoting and quoted content. 
Since tweets were collected based on political keywords, we do not have all the replies to each original post. 
As a result, we lose a substantial portion of the reply chains (i.e., replies to replies).
Therefore, we only retain direct replies to each original post. 
The final dataset consists of \nposts{} original tweets and \nreplies{} direct replies.

We use GPT-4o-mini (gpt-4o-mini-2024-07-18) \citep{gpt4omini} to classify the political leaning of each post as Democratic, Republican, or Unsure, where Unsure captures posts that express no clear partisan stance or fall outside the scope of U.S. electoral politics. 
The system and user prompts are shown in Appendix. 
We validated the classifier on a gold-standard sample of \ngold{} posts, sufficient to obtain a 95\% confidence interval with $\pm 5\%$ margin of error. 
Two coders independently annotated these posts as Democratic, Republican, or Unsure with disagreements resolved through discussion to reach consensus. 
The classifier achieved macro $F1 \approx 0.73$; per-class accuracy metrics are reported in Table~\ref{tab:classifier}.

We classified the political leaning of all reply authors using the same model. We provided the original tweet, the reply text, and the reply author's bio for each reply. The user prompt is shown in Appendix.
We validated the reply classifier on a sample if \ngold{} replies using the same annotation procedure, achieving macro $F1 \approx 0.81$; accuracy metrics are shown in Table~\ref{tab:classifier}.
Since each reply is classified independently, we aggregate across all replies from the same author via majority vote to obtain a stable user-level partisan label.

We assigned a toxicity score to all posts and replies using the Perspective API \citep{lees2022newgenerationperspectiveapi}, which returns a continuous score between 0 and 1, where higher scores indicate greater toxicity. 

Four Mann-Whitney U tests \citep{mannwhitney} were conducted: comparing the toxicity of Democratic versus Republican posts; comparing the toxicity of replies to Democratic versus Republican posts; and comparing the toxicity of Democratic versus Republican replies to Democratic posts, and to Republican posts. 
Reported p-values are calculated using two-sided tests for overall differences, one-sided tests for directional hypotheses. 
We use the Mann-Whitney U test rather than a t-test because toxicity distributions are heavily skewed (see Figure~\ref{fig:tox_dist}): the vast majority of content is low-toxicity, with a small proportion of highly toxic posts, making normality an unreasonable assumption.

All experiments were run on a machine with 32 CPU cores and 115 GB RAM.
Data and code to reproduce our analyses are available at \url{github.com/osome-iu/US2024-election-toxicity}.

\begin{table*}
\centering
\caption{Per-class precision (P), recall (R), and F1 for the post and reply classifiers, validated on a \ngold{}-item sample. Macro-averaged F1 is shown in the bottom row.}
\label{tab:classifier}
\begin{tabular}{lccc|ccc}
\hline
 & \multicolumn{3}{c|}{\textbf{Posts}} & \multicolumn{3}{c}{\textbf{Replies}} \\
\textbf{Class} & \textbf{P} & \textbf{R} & \textbf{F1} & \textbf{P} & \textbf{R} & \textbf{F1} \\
\hline
Democrat   & 0.766 & 0.847 & 0.804 & 0.944 & 0.835 & 0.886 \\
Republican & 0.826 & 0.788 & 0.807 & 0.960 & 0.898 & 0.928 \\
Unsure     & 0.611 & 0.567 & 0.588 & 0.507 & 0.809 & 0.623 \\
\hline
Macro F1   &&& 0.733 &&& 0.812 \\
\hline
\end{tabular}
\end{table*}

\section{Results}

\begin{figure}
\centering
\includegraphics[width=0.9\linewidth]{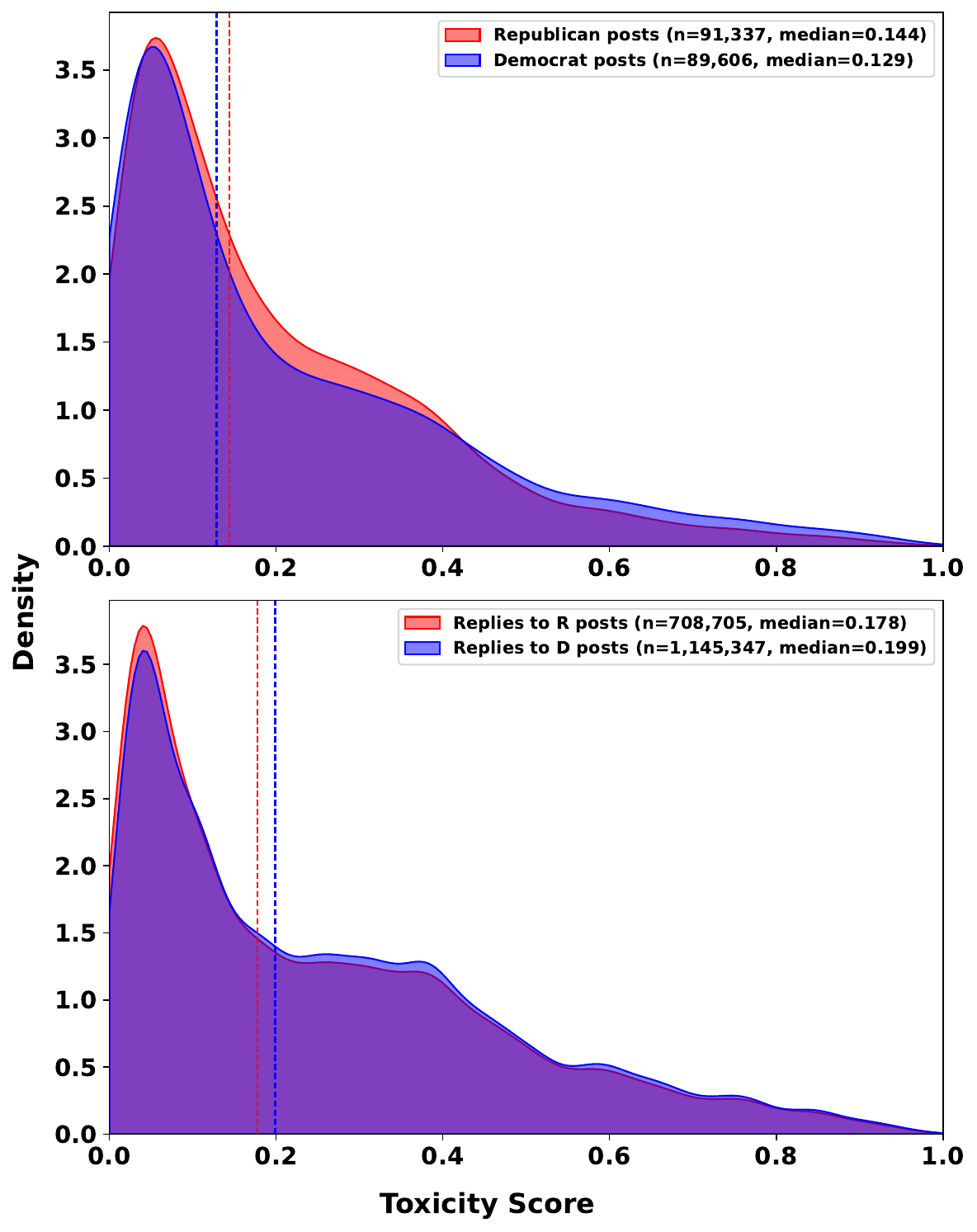}
\caption{Kernel density distributions of post toxicity scores (0--1, higher = more toxic).
Top: Republican-leaning posts (median = 0.144) are more toxic than Democratic-leaning posts (median = 0.129). 
Bottom: replies to Democratic-leaning posts (median = 0.199) are more toxic than replies to Republican-leaning posts (median = 0.178). 
Dashed vertical lines mark group medians. 
Both differences are statistically significant (see Table~\ref{tab:mwu}).}
\label{fig:tox_dist}
\end{figure}

Republican posts are significantly more toxic than Democratic posts ($p < 10^{-6}$; Figure~\ref{fig:tox_dist}, top panel).
Replies tell the opposite story: Democratic posts attract significantly more toxic replies than Republican posts ($p < 10^{-6}$; Figure~\ref{fig:tox_dist}, bottom panel). 

\begin{figure}
\centering
\includegraphics[width=0.9\linewidth]{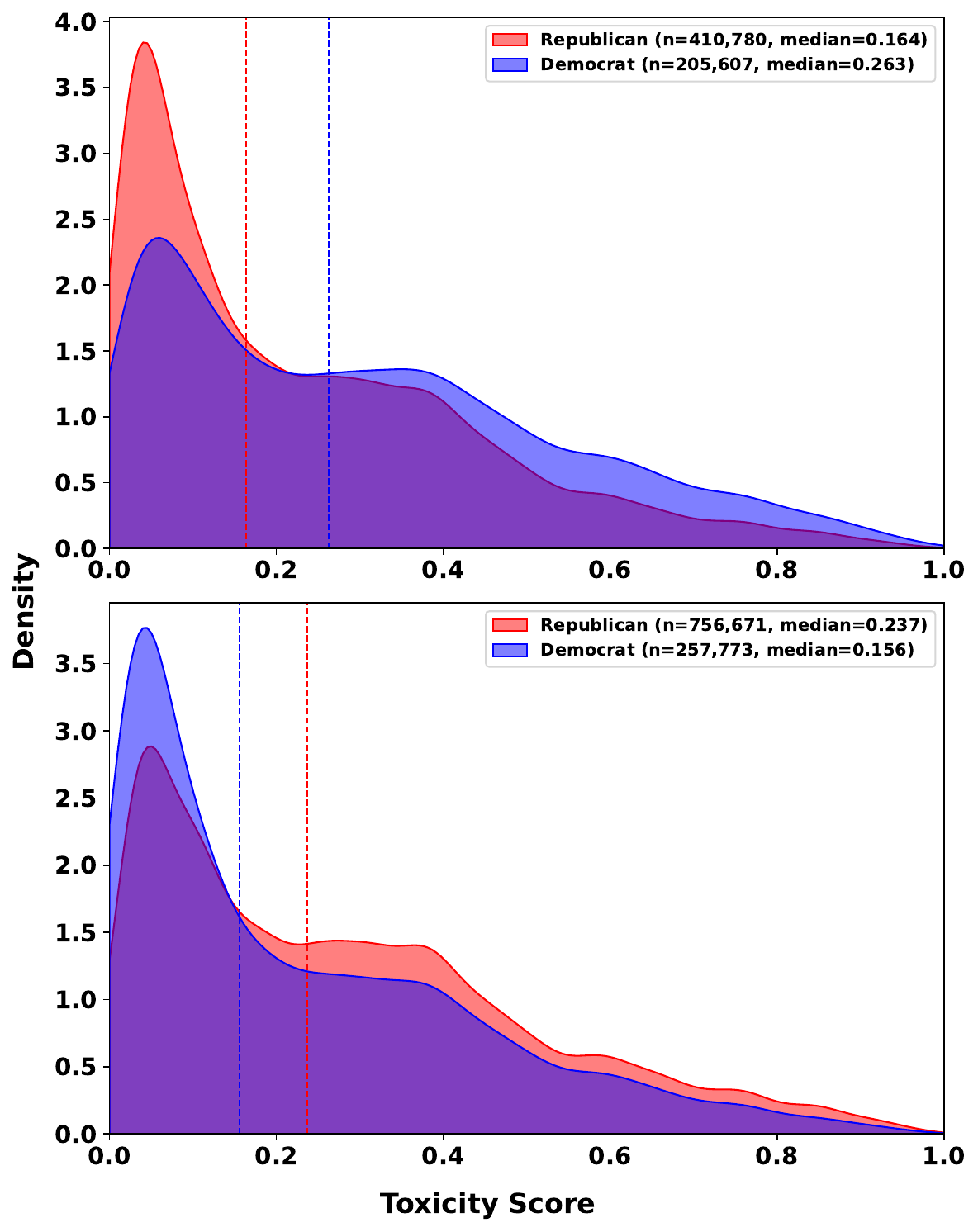}
\caption{Kernel density distributions of reply toxicity scores (0--1, higher = more toxic) by replier partisanship.
Top: among replies to Republican posts, Democratic-leaning replies (median = 0.263) are more toxic than Republican-leaning replies (median = 0.164).
Bottom: among replies to Democratic posts, Republican-leaning replies (median = 0.237) are more toxic than Democratic-leaning replies (median = 0.156). 
In both cases cross-partisan replies are significantly more toxic than same-partisan replies (see Table~\ref{tab:mwu}). 
Dashed vertical lines mark group medians.} 
\label{fig:reply_composition}
\end{figure}

Further analysis is needed to understand how these patterns emerge. 
Cross-partisan replies are more toxic than same-partisan replies in both directions (Figure~\ref{fig:reply_composition}): Republican replies to Democratic posts are more toxic than Democratic replies to Democratic posts, and Democratic replies to Republican posts are more toxic than Republican replies to Republican posts. 
Full statistical results for all comparisons are reported in Table~\ref{tab:mwu}.
Effect sizes are small for cross-partisan comparisons and negligible overall.

However, the volume of cross-partisan replies is widely different. 
Of 1,145,347 replies to Democratic posts, 756,671 (66.06\%) were labelled Republican and 257,773 (22.51\%) were labelled Democrat.
Replies to Republican posts are more balanced: of 708,705 replies, 410,780 (57.96\%) were labelled Republican and 205,607 (29.01\%) were labelled Democrat.
In absolute terms, Republican-labelled accounts sent nearly three times as many cross-partisan replies as Democrat-labelled accounts (756,671 vs.\ 205,607).
At the user level, 269,595 unique Republican accounts engaged with Democratic posts versus 74,081 Democratic accounts engaging with Republican posts. 
The sheer volume of Republican cross-partisan engagement therefore amplifies the toxicity directed at Democratic posts.

\begin{table*}
\centering
\caption{Mann-Whitney U test results for all four toxicity comparisons. $p_{\text{two}}$ = two-sided; $p_{\rightarrow}$ = one-sided in the direction of the stated hypothesis; $p_{\leftarrow}$ = one-sided in the opposite direction. $\delta$ = Cliff's delta (effect size).}
\label{tab:mwu}
\resizebox{\linewidth}{!}{%
\begin{tabular}{lcccc}
\hline
\textbf{Comparison} & $p_{\text{two}}$ & $p_{\rightarrow}$ & $p_{\leftarrow}$ & $\delta$ \\
\hline
R posts more toxic than D posts              & $< 10^{-6}$ & $< 10^{-6}$ & $1.0$ & $0.014$ \\
Replies to D posts more toxic than to R     & $< 10^{-6}$ & $< 10^{-6}$ & $1.0$ & $0.038$ \\
R replies to D posts more toxic than D replies to D posts     & $< 10^{-6}$ & $< 10^{-6}$ & $1.0$ & $0.162$ \\
D replies to R posts more toxic than R replies to R posts     & $< 10^{-6}$ & $< 10^{-6}$ & $1.0$ & $0.193$ \\
\hline
\end{tabular}}
\end{table*}

\section{Discussion}

Our findings reveal a structural asymmetry in how partisan toxicity operates on X during an electoral period: the side that posts more hostile content is not the side that receives the most hostile replies. 
This decoupling of posted and received toxicity suggests that online political hostility is better explained by cross-partisan engagement behavior than by original content.

Our analysis extends prior observations about out-group engagement and cross-partisan toxicity \citep{Xu2024,Rathje2021} by identifying the behavioral source of this asymmetry.
Cross-partisan replies are slightly but significantly more toxic than same-party replies in both directions; what differs is volume: Republicans generate more replies to Democratic content than vice versa.  

The behavioral pattern we observe also speaks to research on cross-partisan exposure and polarization. 
Cross-partisan exposure on social media has been shown to entrench rather than reduce partisan animosity, and limiting such exposure can attenuate it \citep{Bail2018}. 
Our findings show that Republican-leaning accounts engage with Democratic content at disproportionate scale, outnumbering Democratic-leaning accounts who engage in similar cross-partisan activity by a 3.6:1 ratio.
Such a reply asymmetry is consistent with a structural mechanism through which hostility toward Democratic content is amplified.

Several limitations should be noted. 
First, our classifier's lower accuracy on the Unsure category suggests that politically ambiguous content is sometimes assigned a wrong partisan label, which may introduce some noise. 
Second, the Perspective API was trained primarily on Wikipedia and news comments and may not fully capture the informal, ironic, or coded language common on X. 
Third, tweets and replies that lack political keywords are not captured, therefore our analysis is based on incomplete coverage of cross-partisan exchanges.
Fourth, our findings are specific to X during a single electoral cycle; the platform's algorithmic curation and user composition may not generalize to other platforms or non-electoral periods. 
Finally, we cannot establish causality. Republican users disproportionately target Democratic content in volume, but our data does not reveal whether this asymmetry arises from algorithmic exposure, deliberate seeking of opposing content, or coordinated behavior.

Several directions remain for future work.
Temporal analysis around key electoral events---debates, endorsements, election night---would clarify whether cross-partisan targeting spikes episodically or persists as a baseline structural pattern.  
Examining the per-user distribution of cross-partisan replies would reveal whether this behavior is driven by a small, highly active group or spread across the broader user base. 
Network analysis could reveal whether cross-partisan attackers form coordinated clusters or act independently, and whether hostility propagates through retweet cascades. 
Finally, replication on other platforms and in non-U.S. contexts would establish the generality of the asymmetry we observe.

\section*{Acknowledgments}

We would like to thank Emilio Ferrara for organizing a hackathon that inspired this project. 
We also thank the Luddy School of Informatics, Computing, and Engineering at Indiana University for supporting this work, and Jetstream2 \citep{jetstream2} for providing computing resources.

\bibliography{references}

\appendix

\section{Prompt for Posts and Reply Classification}

System prompt to classify posts:

\begin{quote}\texttt{You are a political text analyst.}\end{quote}

\noindent User prompt to classify posts:

\begin{quote}
\texttt{Analyze the following social media post that discusses politics. Determine whether the content of this post more closely aligns with Democratic or Republican political views in the United States. Focus on the political positions, values, or policy preferences expressed in the text. If the post criticizes one political side, it often (but not always) indicates support for the opposite side. If the post is ambiguous or there is not enough information to make a determination, please return ``Unsure.''}
\end{quote}

\noindent User prompt to classify replies:

\begin{quote}
\texttt{%
Given an original tweet, a reply to that tweet, and the bio of the user who wrote the reply, classify the replier as ``Democrat'', ``Republican'', or ``Unsure''.\\[0.5em]
Your task: Infer partisan affiliation from available evidence. Most people in this dataset have political leanings---your job is to detect them.\\[0.5em]
REPUBLICAN indicators:\\
- Support/defense of Trump, Vance, or GOP politicians\\
- Criticism/attacks on Harris, Biden, Walz, or Democrats\\
- MAGA language, \#Trump2024, pro-Trump hashtags\\
- Conservative policy positions (border, guns, taxes, traditional values)\\
- Self-ID as conservative, Republican, MAGA, patriot\\
- Tone: angry at Democrats, defensive of Republicans\\[0.5em]
DEMOCRAT indicators:\\
- Support/defense of Harris, Walz, Biden, or Democratic politicians\\
- Criticism/attacks on Trump, Vance, or Republicans\\
- Criticism of Biden/Democratic establishment from the left (e.g., ``not progressive enough'', ``sellout'', ``corporate Democrat'')\\
- \#HarrisWalz2024, \#VoteBlue, pro-Democratic hashtags\\
- Progressive policy positions (abortion rights, climate, LGBTQ+ rights, gun control)\\
- Self-ID as liberal, progressive, Democrat, leftist\\
- Tone: angry at Republicans, defensive of Democrats\\[0.5em]
WARNING: Do not confuse left-wing criticism of Democrats (still Democrat) with right-wing criticism of Democrats (Republican). Context matters. Progressives criticizing Democrats for not being left enough are still Democrats.\\[0.5em]
UNSURE---use ONLY when:\\
- Non-US politics (UK Labour/Tory, Canadian parties, etc.)\\
- Truly neutral political commentary with no partisan lean\\
- Explicit support for BOTH parties or genuine contradiction\\[0.5em]
IMPORTANT: Even ONE strong indicator is often enough. Criticism of one party usually means support for the other. Sarcasm toward Democrats = likely Republican. Sarcasm toward Republicans = likely Democrat. Make reasonable inferences from tone and context.\\[0.5em]
Return ONLY valid JSON:\\
\{\\
~~~~``classification'': ``Democrat $|$ Republican $|$ Unsure''\\
\}%
}
\end{quote}

\end{document}